\documentclass[usenatbib, useAMS, a4paper]{mn2e}
\usepackage{times}
\usepackage{color}
\usepackage{tensor}
\usepackage{multirow}
\usepackage{graphicx}
\usepackage{amsmath}
\usepackage{natbib}
\usepackage{myaasmacros}
\usepackage[english]{babel}
\usepackage{mathptmx}

\newcommand{\revised}{}
\newcommand{\Msol}{\mathrm{M}_{\odot}}
\newcommand{\kpc}{\mathrm{kpc}}
\def\lsim{\mathrel{\rlap{\lower3pt\hbox{$\sim$}}
    \raise1pt\hbox{$<$}}}                
\def\gsim{\mathrel{\rlap{\lower3pt\hbox{$\sim$}}
    \raise1pt\hbox{$>$}}}                

\begin{document}
\title{How supernova feedback turns dark matter cusps into cores}
\author{Andrew Pontzen}
\author[A. Pontzen and F. Governato]{Andrew Pontzen$^{1,2,3}$\thanks{Email: andrew.pontzen@astro.ox.ac.uk}, Fabio Governato$^{4}$ \\
  $^{1}$Kavli Institute for Cosmology and Institute of Astronomy, Madingley Road, Cambridge CB3 0HA, UK \\
  $^{2}$Emmanuel College, St Andrew's Street, Cambridge, CB2 3AP UK \\
  $^{3}$Oxford Astrophysics, Denys Wilkinson Building, Keble Road, Oxford
  OX1 3RH, UK \\
  $^{4}$ Astronomy Department, University of Washington, Seattle, WA 98195, USA}

\date{Accepted ---. Received ---; in original form ---}
\maketitle

\newcommand{\pc}{\mathrm{pc}}
\newcommand{\cm}{\mathrm{cm}}
\newcommand{\Myr}{\mathrm{Myr}}
\newcommand{\dd}{\mathrm{d}}
\newcommand{\Gyr}{\mathrm{Gyr}}
\newcommand{\ergs}{\mathrm{ergs}}
\begin{abstract}
  We propose and successfully test against new cosmological
  simulations a novel analytical description of the physical processes
  associated with the origin of cored dark matter density
  profiles.  In the simulations, the potential in the central
  kiloparsec changes on sub-dynamical timescales over
  the redshift interval $4>z>2$ as repeated, energetic feedback
  generates large underdense bubbles of expanding gas from
  centrally-concentrated bursts of star formation. The model
  demonstrates how fluctuations in the central potential irreversibly
  transfer energy into collisionless particles, thus generating a dark matter
  core.  A supply of gas undergoing collapse and rapid expansion is
  therefore the essential ingredient.  The framework, based on a novel
  impulsive approximation, breaks with the reliance on adiabatic
  approximations which are inappropriate in the rapidly-changing
  limit.  It shows that both outflows and galactic fountains can give
  rise to cusp-flattening, even when only a few per cent of the
  baryons form stars. Dwarf galaxies maintain their core to the
  present time.  The model suggests that constant density dark matter
  cores will be generated in systems of a wide mass range if central
  starbursts or AGN phases are sufficiently frequent and energetic.
\end{abstract}

\section{Introduction}

Over the last two decades, galaxies formed in numerical simulations
based on the inflationary $\Lambda$CDM paradigm have suffered from a
number of well-documented mismatches with observed systems.  One of
the most prominent of these has been the rotation curves of
disk-dominated dwarf galaxies (e.g. \citeauthor{1994Natur.370..629M}
1994; \citeauthor{flores94} 1994; for more recent updates see
\citeauthor{blitz05} 2005, \citeauthor{2010arXiv1011.0899O}
\citeyear{2010arXiv1011.0899O} and references therein). The observed
kinematics imply a constant density core of dark matter interior to
1~kpc, whereas simple physical arguments and simulations suggest that
the cold dark matter density should be increasing roughly as $\rho
\propto r^{-1}$ to vastly smaller radii
\cite[e.g.][]{1991ApJ...378..496D,1996ApJ...462..563N}.

Since some of the earliest work on dark matter profiles it has been
suggested that sufficiently violent baryonic processes might be
responsible for heating dark matter cusps into cores
\citep{flores94}. The proposed mechanisms identified in these papers
fall in two broad categories: supernova-driven flattening
\citep{1996MNRAS.283L..72N,gelato99,2001MNRAS.321..471B,2002MNRAS.333..299G,mo04,2005MNRAS.356..107R,2006Natur.442..539M,2008Sci...319..174M},
and dynamical friction from infalling baryonic clumps or disk
instablities
\citep{2001ApJ...560..636E,2002ApJ...580..627W,2006ApJ...649..591T,2008ApJ...685L.105R,2009ApJ...702.1250R,2010A&A...514A..47P,2010ApJ...725.1707G,2011arXiv1105.4050C}. Within
the former category, most early works focused on a single, explosive
mass-loss event. It then became clear that even with extreme
parameters, such an event transferred insufficient energy to dark
matter particles \citep{2002MNRAS.333..299G}. On the other hand,
\cite{2005MNRAS.356..107R} showed that several more moderately violent
bursts could be effective in creating a core.  Increasingly
sophisticated numerical work by
\cite{2006Natur.442..539M,2008Sci...319..174M} strongly supported the
notion of stellar feedback and energy transfer from baryons to dark
matter as the generator of cores, but did not fully explain the
physical mechanism behind this transfer or follow the evolution of
dwarf galaxies to $z=0$ to ensure that the cores were long-lived.

Recently, simulations were able to produce realistic, present-day
cored dwarf galaxies within a fully cosmological context
\citep[][henceforth G10]{2010Natur.463..203G}. These simulations
resolve individual star formation `clumps' at the density of molecular
clouds leading to galaxies that are additionally realistic because,
like many observed dwarfs \citep{2009MNRAS.396..121D}, they have no
bulge -- a consequence of preferentially expelling
low-angular-momentum gas from the progenitors via naturally occurring
galactic winds (\citeauthor{2010arXiv1010.1004B} 2011; see also
\citeauthor{2001ApJ...555..240B} 2001 and
\citeauthor{2001MNRAS.326.1205V} 2001).  \cite{2010arXiv1011.2777O}
confirmed that these effects bring the simulations into excellent
agreement with observational constraints on stellar and H{\sc i}
content as well as those on overall mass distribution. By testing
against dark-matter-only runs, G10 provided strong support to a model
where the core flattening is generated by baryonic effects, in
particular by rapid gas motions; moreover, a suite of comparison
simulations revealed that these effects only become significant if
stars form in dense clumps ($\sim 100$ amu $\mathrm{cm}^{-3}$), suggesting that energy injection has to be
concentrated in local patches \cite[see also][]{2007arXiv0712.3285C}.

The comparison of simulations with recent
  observations \citep{2010arXiv1011.2777O} highlighted that
  feedback must occur in numerous relatively mild events to allow thin disks to form. 
  However given the lack of analytic framework for understanding the
  microphysics of this process, the precise mechanism of
  supernova-driven cusp flattening was not further elucidated in G10.
  Providing such a framework is the aim of the present work.

  The remainder of this paper describes how to model the effects of
  small, central starbursts which create pockets of rapidly expanding
  gas and strong fluctuations in the local potential. Over time these
  repeated processes gradually transfer energy from the gas to the
  dark matter component.  This has much in common with the view of
  \cite{2006Natur.442..539M,2008Sci...319..174M} but places more
  emphasis on disrupting clumps as opposed to pushing them around, and
  clarifies that resonance\footnote{Although
    \cite{2008Sci...319..174M} described their model as `resonant',
    they have since stated that they did not mean to invoke a
    formal resonance, but rather the notion of changes in the
    potential occurring on roughly the dynamical timescale
    (Wadsley, pri. comm.).}  is not required.  Because our picture can
  be modelled mathematically, we are able to validate it against the
  simulations, showing that the envisioned process indeed creates
  cores within the G10 simulations. 

  This paper is organized as follows. Section~\ref{sec:first-sims}
  introduces improved simulations based on those of G10, and discusses
  the characteristics of these simulations which predict
  cusp-flattening, thus motivating a study of orbits in rapidly
  changing potentials (Section~\ref{sec:virial-eqs}).  The initial
  discussion is, for simplicity, limited to power-law potentials but
  Section~\ref{sec:second-sims} removes this restriction, presenting
  more general equations to explain the detailed simulation
  results. We relate our work to the wider literature and conclude in
  Section~\ref{sec:conclusions}, {\revised also discussing the realism
    of the underlying hydrodynamical evolution within the simulated
    galaxies.} In a companion paper (Governato et al, in prep) we will
  discuss the scaling of the dark matter cores with galaxy masses.

\section{The simulations}\label{sec:first-sims}

The smoothed particle hydrodynamics (SPH) simulations, run using the
{\it Gasoline} code \citep{2004NewA....9..137W}, are closely related
to and improve upon those described in more detail by G10. {\revised
  The emphasis of the present work is on interpreting dynamical
  effects, rather than discussing the numerical methods in depth;
  however see Section \ref{sec:conclusions} for brief comments on
  computational accuracy.} Our new runs output more regular timesteps
and include the effects of metal-line cooling according to the
prescription of \cite{2010MNRAS.407.1581S}. The simulations in this
paper focus on the region hosting the galaxy denoted `DG1' in G10. The
`zoom' technique \cite[e.g.][]{1993ApJ...412..455K} allows for a high
mass resolution of $M_{\mathrm{p}} = 3 \times 10^3 \, \Msol$ (for gas
particles) and $M_{\mathrm{p}} = 1.6 \times 10^4\,\Msol$ (dark matter)
with a softening of $86\,\pc$ in a full $\Lambda$CDM cosmological
context.  We conducted analysis on the two most massive systems within
this region: DG1 itself ($M_{\mathrm{vir}}=3.7 \times 10^{10}\,\Msol$
at $z=0$) and a somewhat smaller galaxy ($M_{\mathrm{vir}}=1.3 \times
10^{10}\,\Msol$ at $z=0$). Most results will be presented for the
latter case, because the former undergoes a major merger at
$z=3$. Although our model does predict the correct flattening for DG1,
its volatile merger history would introduce undue complexities into
our discussion.

In the first run, denoted HT (``high threshold''), stars are allowed
to form only at hydrogen densities exceeding $100\,\mathrm{cm}^{-3}$.
The second run, LT (``low threshold''), is identical to the first
except that it allows stars to form at densities exceeding
$0.1\,\mathrm{cm}^{-3}$.

{\revised Adopting the higher threshold for star formation is strongly
  motivated by observational evidence that molecular clouds form at
  such densities \citep{2008AJ....136.2846B,2010Natur.463..781T}.
  HT thus exhibits more realistic behaviour of the interstellar medium
  \cite[e.g.][]{2008arXiv0802.0961S}. In particular, only at high
  formation thresholds exceeding $\sim 10\,\cm^{-3}$ will supernova
  feedback naturally give rise to bulk gas motions and outflows
  \citep{2007arXiv0712.3285C}. This follows because individual
  high-density clumps are efficient in converting gas to stars,
  ultimately leading to vast overpressurization of the clump from the
  high local density of supernovae. The particular threshold value of
  $100\,\cm^{-3}$ for HT is chosen for consistency with G10, in which
  it was argued that this is the highest density which can be
  considered physical at our present resolution. We have also verified
  that when H$_2$ physics is consistently included in simulations
  (Christensen et al, submitted; Governato et al, submitted) star
  formation indeed proceeds only at high densities even in the absence
  of an explicit threshold.

  By contrast, LT's threshold of $0.1\,\cm^{-3}$ represents an
  approximate historical norm for galaxy formation simulations
  \cite[e.g.][]{1993MNRAS.265..271N,1996ApJS..105...19K} broadly
  compatible with the observed cut-off in star formation at low column
  densities averaged over $\sim\kpc$ scales
  \cite[e.g.][]{2007ApJ...671..333K}. Until recently LT would have
  been the most motivated choice; however, with the addition of
  metal-line cooling at increasing resolution we now
  prefer the HT simulations, using LT as a reference to understand why
  older simulations did not produce the effects of interest here.}

 As expected following G10, LT remains cusped, unlike HT
which develops a $1\,\kpc$ dark matter core at $z=0$ in both of its
two most massive halos.  In all cases the code consistently follows
the feedback effects of the stellar populations
\citep{2006astro.ph..2350S} so that, after a delay of $\sim 10 \,
\Myr$, significant amounts of thermal energy are deposited into the
surrounding gas. By $z=2$, when HT has developed a stable core, LT and
HT runs have formed an almost identical mass of stars ($7 \times
10^{7} \Msol$) and therefore the same quantity of supernova energy has
been released ($7 \times 10^{56}\, \ergs$).\footnote{Note, however,
  that by $z=0$ the LT simulation forms $4 \times 10^9 \,\Msol$ in
  stars compared against the HT simulation's $5 \times 10^8\, \Msol$.
  Only the HT simulation forms a realistic dwarf galaxy
  \protect\citep{2010arXiv1011.2777O}. } The failure of LT to lose its cusp
thus reflects a difference in the coupling mechanism, not in the
absolute energy deposition.

\begin{figure}
\includegraphics[width=0.5\textwidth]{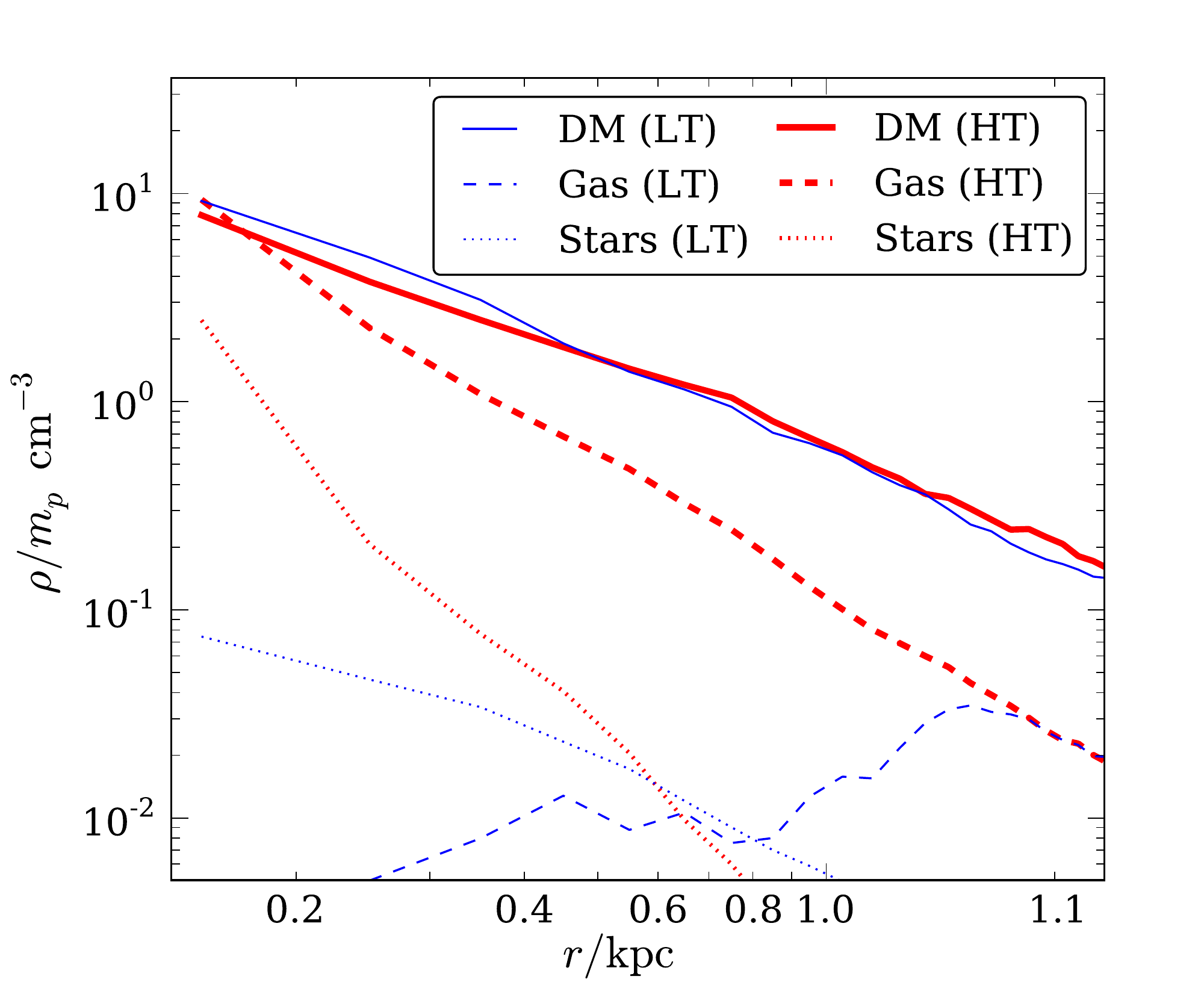}
\caption{Spherically averaged halo density profiles for high star-formation
  threshold (HT, thick red lines) and low threshold (LT, thin blue lines)
  simulations at $z=4$, shortly before the dark matter profile starts
  to flatten in the high-threshold model. Solid, dashed and dotted
  lines show respectively dark matter, gas and stellar content.  In
  the low threshold case, dark matter dominates by orders of magnitude
  at every radius. In the high threshold case, the gas reaches a
  comparable density to the collisionless matter in the central
  regions.  Gaseous processes can therefore cause heating of
  collisionless components (dark matter and stars) in HT but not LT
  runs.}\label{fig:den}
\end{figure}

Figure~\ref{fig:den} gives immediate insight into the difference
between HT (cusp-flattening) and LT (cusp-preserving) simulations by
showing their spherically-averaged halo density profiles shortly
before the cusp begins to flatten in HT, at $z=4$. Solid, dashed and
dotted lines indicate respectively dark matter, gas and stellar
density; thick red and thin blue lines represent the HT and LT runs in
turn. In the LT run, the gas density cannot exceed the threshold of
$0.1\,\mathrm{cm}^{-3}$: stars are able to form and deposit supernova
energy, preventing further cooling.  In the HT run, by contrast, the
gas density rises monotonically towards the centre because most of the
gas is not eligible to form stars. The result is that the central gas
density slightly exceeds that of the dark matter. It is natural to
suppose from this that the halo will have a qualitatively different
reaction to the presence of baryons in the two runs. (This work
focuses on expansion of dark matter orbits, but we verified that the
same processes operate on the similarly collisionless star particles
in the simulation, the interesting implications of which are left for
future study.)

\begin{figure}
\includegraphics[width=0.5\textwidth]{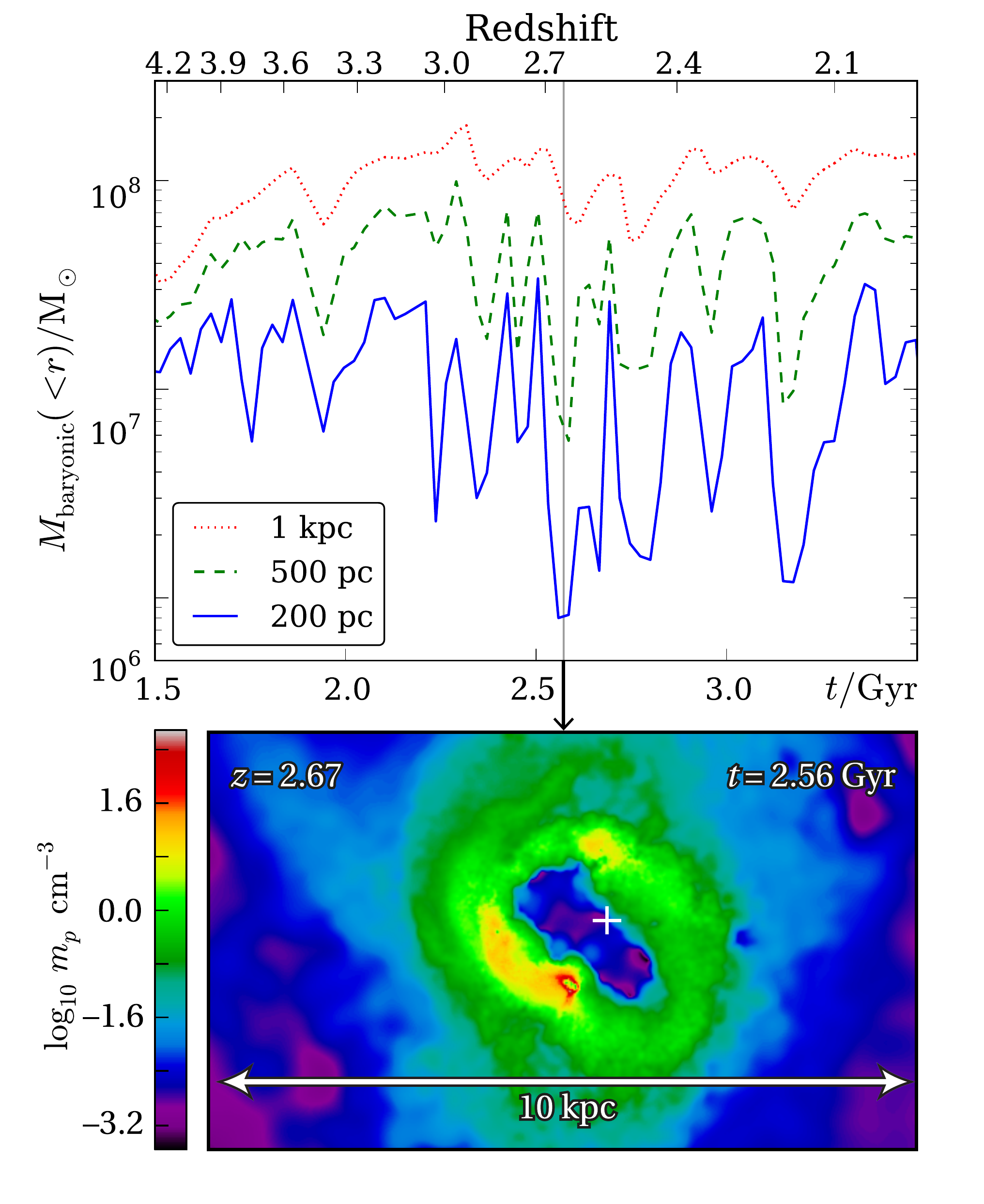}
\caption{(Upper panel) The baryonic mass interior to, from top line to
  bottom, $1\, \kpc$, $500\, \pc$ and $200\, \pc$ (HT
  simulation). Bursty central star formation coupled to strong
  supernova feedback causes coherent, rapid oscillations in the
  potential interior to $1\, \kpc$. The orbital time of typical dark
  matter particles interior to $1\, \kpc$ is $\gsim 25\, \Myr$.  By
  contrast the simulated supernova bubbles can encompass the inner
  kiloparsec in around $3\,\Myr$, far too rapidly for the adiabatic
  approximation to be valid. The lower panel shows the disk-plane
  density during the starburst event at $t=2.56\,\Gyr$, $z=2.67$. A
  large underdense bubble has formed at the centre of the disk through
  thermal expansion of gas heated by multiple supernova
  explosions. The cross marks the halo
  centre. }\label{fig:HT-fluctuations}
\end{figure}

Focusing on the HT simulation, Figure~\ref{fig:HT-fluctuations} (upper
panel) shows the baryonic mass enclosed within $0.2$, $0.5$ and
$1.0\,\kpc$ as a function of time. From around $1.7\,\Gyr$ after the
big bang, the density near the centre undergoes order-of-magnitude
fluctuations. First, gas flows in, cools and condenses near the centre
of the potential well. Then, as the density of clumps rises above the
$100\,\cm^{-3}$ threshold, star formation is allowed to proceed.

Once supernova energy is dumped into the gas, thermal expansion forms
an underdense bubble of up to several kiloparsecs diameter (lower
panel, Figure~\ref{fig:HT-fluctuations}).  The energy is initially
deposited within a small volume (a consequence of the high star
formation threshold) which then reaches temperatures of $\sim
10^8\,\mathrm{K}$. The gas is vastly overpressurized relative to its
surroundings, so expands at close to the thermal velocity (typically
reaching $\sim 300\,\mathrm{km}/\mathrm{s} \simeq 0.3
\,\mathrm{kpc}/\mathrm{Myr}$). Compared to the orbital
timescales, which are $\gsim 25 \, \mathrm{Myr}$, the bubble
formation is effectively instantaneous.

The adiabatic cooling from the expansion is included, but the bubbles
nonetheless remain hot ($\sim 10^{6}\,\mathrm{K}$) by the time they
reach rough pressure equilibrium with the remainder of the disk. They
are then sufficiently underdense ($\sim 10^{-2}\,\cm^{-3}$) that the
radiative cooling timescale is up to $100\,\Myr$; the bubble can
therefore persist for this length of time, after which it cools back
into the disk if it has not actually escaped in a galactic wind.  The
rapid, repeated fluctuations in the central mass content of the
simulated galaxy are similar to those shown in Figure 3 of
\cite{2008Sci...319..174M}. However we have verified that, in our
case, this is due to the gas being heated and expanding outward rather
than remaining in rapidly-moving coherent clumps as suggested by
\cite{2008Sci...319..174M}.

The overall picture for the dark matter is insensitive to the ultimate
destination of the gas, requiring only the intermittent variations
explained above; the present work makes no assumptions about mass
loss. Note, however, that significant winds do exist in the
simulations; the final baryon fraction in the galaxies is only $25\%$
of the cosmic value (G10) and only 3\% of the baryons have been turned
into stars.  The winds have been shown to be important in matching the
star formation rates, distribution of stars and final baryon fractions
of the dwarf galaxies
\citep[see][]{2010ApJ...708L..14M,2010arXiv1010.1004B,2010arXiv1011.2777O}.

\section{Analytical model}\label{sec:virial-eqs}

This Section discusses how the energy of a single dark matter
particle (or star) changes in response to a fluctuating 
potential sourced by gas subject to processes described above.  Two
restrictions on the calculation are imposed throughout the paper:
\begin{enumerate}
\item the potential is assumed to be spherically symmetric;
\item the tracer particles are assumed to be massless, {\it i.e.}
  the potential is always external.
\end{enumerate}
The latter condition represents a decision to focus on the
microphysical mechanism via which particles gain energy, rather than
the subsequent evolution of the self-gravitating system. The first
condition could in future be relaxed, but makes calculations much
simpler because particles orbit in the 1D effective potential
\begin{equation}
V_{\mathrm{eff}}(r;j,t) = V(r; t) + \frac{j^2}{2r^2}\textrm{,}\label{eq:veff}
\end{equation}
where $V(r;t)$ is the time-dependent physical potential and $j$ is a
conserved angular momentum. 
 
For simplicity we will temporarily impose two further restrictions
which will later be removed in Section \ref{sec:second-sims}:
\begin{enumerate}
\setcounter{enumi}{2}
\item only the normalization of the potential changes, {\it i.e.} its
functional form is fixed;
\item the functional form of the potential is a power law.
\end{enumerate}
Together these imply that the underlying potential in~(\ref{eq:veff})
is specified by $V(r; t) = V_0(t) r^n$. Some useful test cases fall
into this exact form: a Keplerian orbit has $n=-1$ while
a harmonic oscillator implies $n=2$. The final case will be of
particular interest chiefly for its analytic simplicity, but we also
note that it corresponds to assuming a spatially constant density of
matter.

The rate of change of the total energy of a particle orbiting
within the potential, $\dd E/\dd t$, is given by the partial
derivative $\left. \partial V/\partial t \right|_{r(t)}$, where $r(t)$
denotes the solution to the equations of motion. In the limit of an
instantaneous change in the potential $V \to V+\Delta V$ occurring at
time $t$, the total energy of the particle therefore changes by
$\Delta E = \Delta V(r(t)) = \Delta V_0 \, r(t)^n$.

Assuming we have no prior knowledge of the phase of the particle, the
virial theorem \cite[e.g.][]{goldstein2002classical} states that the
expected potential energy is
\begin{equation}
\langle V \rangle = \frac{2E_0}{2+n}\textrm{,}
\end{equation}
where $E_0$ is the total energy of the particle and the result is
independent of $j$. This and following equations are also therefore
valid in the one-dimensional ($j=0$) subcase.

If suddenly $V_0 \to V_0 + \Delta V_0$, the energy after the potential
change is $E_0 + \Delta E_1$, where
\begin{equation}
  \langle \Delta E_1 \rangle = \Delta V_0 \langle r^n \rangle = \frac{2 E_0}{2+n} \frac{\Delta V_0}{V_0}\textrm{.}\label{eq:deltaE-virial}
\end{equation}
The fiducial adiabatic limit can be attained from here by assuming
$\Delta V_0$ and $\Delta E_1$ to be infinitesimal and integrating
over a series of such changes taking $V_0$ smoothly to $V_f$. This
yields a final, finite change in energy:
\begin{equation}
E_{f,\mathrm{adiabatic}} = E_0\, \left(\frac{V_f}{V_0}\right)^{2/(2+n)}\textrm{.}\label{eq:adiabatic-Efinal}
\end{equation} 
As expected in the adiabatic limit, equation
(\ref{eq:adiabatic-Efinal}) implies no energy shift, regardless of the
intermediate states, if the final potential is the same as the
initial. This is the central problem of the adiabatic
approximation. Figure \ref{fig:HT-fluctuations} shows that the final
distribution of gas will indeed be very similar to the initial, and
therefore that the adiabatic prediction will be for no change in the final
distribution of dark matter.

However Section \ref{sec:first-sims} showed that potential changes in
the simulations take place on timescales much shorter than the
dynamical time, because the expansion speeds of the supernova-induced
bubbles are much larger than the local circular velocity.  As the
  gas expands and leaves the galaxy centre, the potential undergoes a
series of large, instantaneous jumps, invalidating the adiabatic
result given by equation~(\ref{eq:adiabatic-Efinal}).

Instead of integrating, one should therefore recursively apply equation
(\ref{eq:deltaE-virial}) to each finite change.  If, for instance, the
potential switches immediately back to its original depth, the second
shift in energy is given by
\begin{equation}
\langle \Delta E_2 \rangle = \frac{2\left(E_0 + \langle \Delta E_1 \rangle\right)}{2+n} \frac{-\Delta V_0}{V_0 + \Delta V_0}\textrm{,}
\end{equation}
where the angular brackets indicate averaging over the orbital phase
of our chosen trajectory during both the initial and final
instantaneous potential jumps. These conditions are justified because
the initial blowout is not causally connected to the location of a
single tracer particle, nor is the exact fractional number of orbits
between initial blowout and eventual recollapse predictable (this
aperiodicity is illustrated by Figure \ref{fig:HT-fluctuations}).  By
Taylor expanding, we find that the expected final energy of the orbit
is given by
\begin{equation}
\langle E_f \rangle = E_0 + \langle \Delta E_1 + \Delta E_2 \rangle \simeq E_0  + \left ( \frac{\Delta V_0}{V_0} \right)^2 \frac{2n}{(2+n)^2} E_0\,\textrm{,}\label{eq:power-potential-energy-change}
\end{equation}
which is always an energy gain for bound orbits since $E_0<0$ for
$n<0$. The energy gain is second order  in the potential change
$\Delta V_0$, but linear in the energy. 
One may verify that, if the potentially first changes suddenly but then
gradually (i.e. adiabatically) relaxes to its original state, we will
also see an increase in expected final energy of the same magnitude.
The essential point is for the initial change to be rapid; the
energy shift will then follow.



The special case of the harmonic oscillator ($n=2$) is 
helpful in demonstrating the origin of this energy shift, because its
dynamics are especially simple. The potential is separable in
Cartesian coordinates which means that we can assume the
one-dimensional subcase without loss of generality. Then, for the
case of sudden discrete jumps, the analytic form of the solution
is written
\begin{equation}
x(t) = A \cos \left(\omega t + \psi \right)\textrm{,}
\end{equation}
where $\omega^2(t)=2 V_0(t)$ while $A(t)$ and $\psi(t)$ specify the
amplitude and phase of the oscillation, which change discontinuously
with the potential. The new values of $A$ and $\psi$ after any jump
can be determined either through energy arguments as above or by
requiring continuity of both $x$ and $\dot{x}$. Without loss of
generality, we let $\omega$ change from $\omega_0$ to $\omega_1$ at
$t=0$. The amplitude of the trajectory after the jump, $A_1$, is given
by the expression
\begin{align}
A_1^2 & = A_0^2 \left[ 1  + \frac{\omega_0^2-\omega_1^2}{\omega_1^2} \sin^2 \psi_0 \right]\textrm{,}\label{eq:harm-amplitude-jump}
\end{align}
which is explicitly dependent on the orbital phase $\psi_0$ at which
the discontinuity occurs. The corresponding change of energy is
\begin{equation}
\Delta E_1 = - E_0 \frac{\omega_0^2-\omega_1^2}{\omega_0^2} \sin^2 \psi_0\textrm{.} \label{eq:harm-energy-jump}
\end{equation}

\begin{figure}
\includegraphics[width=0.5\textwidth]{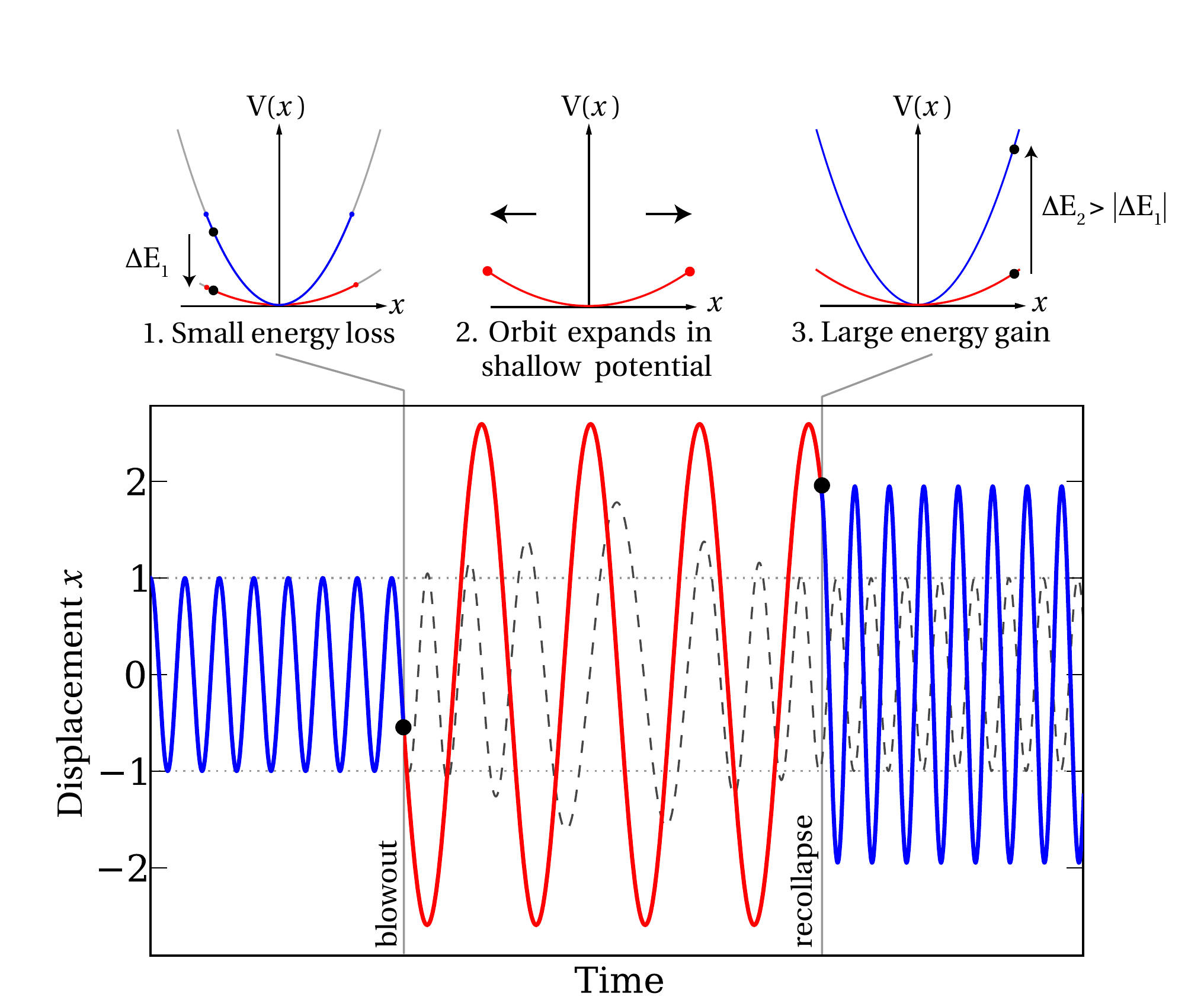}
\caption{The mechanism for injecting energy into the dark matter
  orbits, illustrated by the exact solution for a time-varying
  harmonic oscillator potential.  The lower panel shows (solid line) a
  solution to the equations of motion where $\omega^2=1$ (blue) at
  early and late times, while at intermediate times $\omega^2 = 0.1$
  (red) mimicking baryonic blowout and recondensation. The changes in
  potential occur instantaneously; in this case the final amplitude of
  the oscillation is approximately twice that of the initial orbit.
  The dashed line shows the solution when the potential changes
  smoothly over several orbital periods; this gives adiabatic
  behaviour, so that the final orbit regains the initial amplitude,
  demonstrating the necessity for relatively sudden potential
  jumps. The inset figures (top) illustrate how the post-blowout orbit
  expansion implies that the late-time energy gain  dominates
  over the initial energy loss.
}\label{fig:harmonic-oscillator}
\end{figure}

We can now analyze the changes in a single trajectory which lead us to
recover the expected gain in orbital energy under a single
blowout-recollapse cycle, equation
(\ref{eq:power-potential-energy-change}).
Figure~\ref{fig:harmonic-oscillator} (thick line) shows an example
orbit for which $\omega_0^2 = 10\, \omega_1^2$.  The initial amplitude
is unity until, at a certain time, the potential flattens out as mass
is lost from inside the orbit. Intuitively, or from equation
(\ref{eq:harm-energy-jump}), this must always involve a loss of
energy for the particle (since $\omega_0^2>\omega_1^2$); see also
panel~1 at the top of Figure \ref{fig:harmonic-oscillator}.  However,
according to equation~(\ref{eq:harm-amplitude-jump}), the new
potential is sufficiently flattened that the orbital amplitude is now
larger than unity (panel 2). This is crucial because the energy gain
made when the potential change is later reversed ($\omega$ returns
from $\omega_1$ to $\omega_0$) will be accordingly larger, scaling
with $\langle x^2 \rangle$ (panel 3). Applying equations
(\ref{eq:harm-amplitude-jump}) and (\ref{eq:harm-energy-jump}) for the
reverse jump (i.e. with $\omega_1$ and $\omega_0$ exchanged) allow
this picture to be directly verified.

This makes more concrete the assertion of equation
(\ref{eq:power-potential-energy-change}) that one always expects
to gain energy during a series of potential changes. Yet seen
from another perspective, such a claim still needs
to be reconciled with the underlying dynamics which are fully
time-reversible. For instance, the time-reverse of
Figure~\ref{fig:harmonic-oscillator} (viewing the figure from right to
left) represents an equally valid trajectory of the forwards dynamics
and yet {\it loses} energy.

Under time-reversal, the blowout phase maps onto the recollapse phase
and vice-versa. The irreversibility thus arises not from dynamical
differences but statistical differences between the forwards-blowout
and reverse-recollapse pictures.  In particular a uniform prior on the
orbital phase before a transition is always assumed, $2 \pi\,
p(\psi_0) = 1$. On the other hand the phase $\psi_0'$ after the
transition is determined by
\begin{equation}
\tan \psi_0' = \frac{\omega_0}{\omega_1}\,\tan\psi_0\textrm{,}\label{eq:phase-relation}
\end{equation}
and, accordingly, the probability distribution function of $\psi_0'$ is
\begin{equation}
2 \pi \,p(\psi_0') = \left(\frac{\omega_0}{\omega_1}\cos^2 \psi_0' + \frac{\omega_1}{\omega_0
}\sin^2 \psi_0'\right)^{-1}\textrm{.}\label{eq:psi1-distrib}
\end{equation}
The precise functional form~(\ref{eq:psi1-distrib}) is not crucial,
only that $p(\psi_0')$ cannot be taken to be uniform.  After the
sudden baryonic blowout, collisionless particles enter their new orbit in a special
phase -- preferentially near pericentre -- so that they subsequently
migrate outwards in unison.

It is this difference in knowledge of phases before and after sudden
changes that allows irreversibility in the real universe to appear in
the model. Only if all collisionless particles were near their pericentre just
before the baryons returned would the statistical properties of the
reversed picture match those of the actual model.  While this is
dynamically possible, it is statistically unlikely.

Finally note that if the changes in potential are introduced gradually,
the process should become adiabatic and hence reversible.
The dashed line in Figure \ref{fig:harmonic-oscillator}
shows a numerical solution for which $\omega$ changes smoothly over
several orbital times from $\omega_0$ to $\omega_1$, then back to
$\omega_0$. As expected from equation (\ref{eq:adiabatic-Efinal}), the
final orbital amplitude is the same as its initial value, confirming
the qualitatively different results to be expected from gradual
variation as opposed to sudden jumps.

\section{Validating the analytic model against simulations}\label{sec:second-sims}

To test the picture expounded above we start by generating a
time-dependent effective toy potential from the simulations (Section
\ref{sec:first-sims}). This is given by equation (\ref{eq:veff}), with
$V(r;t)$ calculated from the spherically averaged density profile.
The starting energy $E_0$ and the value of $j$ can be determined by
specifying initial orbital parameters. The angular momentum is
necessarily conserved because of the spherical symmetry of the
modelling (restriction~1 of Section~\ref{sec:virial-eqs}). In the
simulations the changes in potential are not exactly symmetric
(e.g. lower panel of Figure~\ref{fig:HT-fluctuations}); however we
will see below that, for the purposes of calculating real-space
density profiles, the symmetric approximation which enforces constant
$j$ actually works extremely well.

As before, the energy shift for one jump is given by averaging
over possible orbital phases. However the potential
$V_{\mathrm{sphere}}$ is no longer an exact power law, so the
calculation required is
\begin{eqnarray}
\langle\Delta E\rangle & = & \frac{\int \Delta V_{\textrm{eff}}(r(t)) \dd t}{\int \dd t}  \nonumber
\\ & = & \left. {\int \frac{\Delta V_{\mathrm{eff}}(r)\, \dd r}{\sqrt{E-V_{\mathrm{eff}}(r)}}} \right/ {\int \frac{\dd r}{\sqrt{E-V_{\mathrm{eff}}(r)}}}\textrm{,}\label{eq:delta-E1}
\end{eqnarray}
where the time integrals are evaluated over an orbital period; after
changing variables to $r$ this corresponds to integrating over the
region where the integrand is real. Equation~(\ref{eq:delta-E1})
agrees with equation~(\ref{eq:deltaE-virial}) for the special
case of power-law potentials.

The remainder of this Section applies expression~(\ref{eq:delta-E1})
recursively to a time-series of potentials from the HT
(cusp-flattening) simulation, at each step updating $\Delta V$, $E$
and $V_{\mathrm{eff}}$ appropriately\footnote{Because $E$ takes a
  random walk, a more accurate result is in principle attainable by
  keeping track of its evolving distribution function rather than just
  its expected value.  Our approach here is akin to taking the first
  term in a Fokker-Planck analysis and will be an excellent
  approximation because the energy shifts are approximately linear, as
  can be shown by generalizing equation (\ref{eq:deltaE-virial}).}.
The energy gain is evaluated at every stored simulation timestep; the
relevant outputs are written every $\delta t\simeq 27\,\Myr$. Thus
changes occurring on timescales $\le \delta t$ will implicitly be
classified as ``rapid'' (composed of one jump) whereas those occurring
on timescales $\gg \delta t$ will automatically be treated as
``adiabatic'' (composed of many small steps). While the boundary
between these limits cannot be uniquely defined, the change in
behaviour must occur at around the orbital period for a particle,
which is indeed $\sim 25\, \Myr$. We verified by running checks with
only every second timestep ($\delta t \simeq 54\,\Myr$) that the
results presented are insensitive to the precise time-slicing.

\begin{figure}
\includegraphics[width=0.5\textwidth]{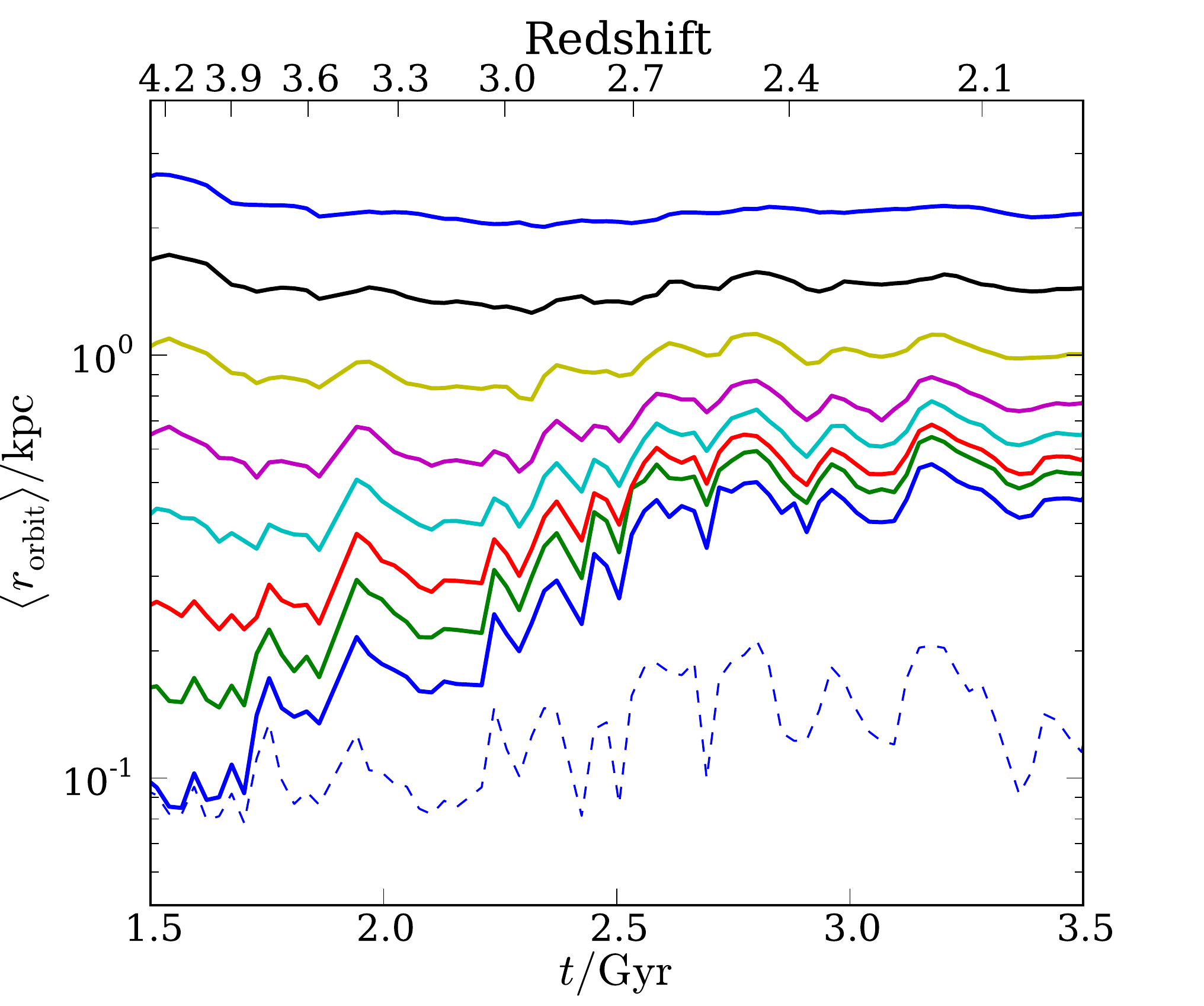}
\caption{Using the spherically averaged potential from the
  simulations, we model the expansion of orbits of test particles at
  different initial radii (solid lines). Orbits starting significantly
  within the inner kiloparsec migrate outwards over several gigayears,
  whereas those starting outside a kiloparsec do not feel the rapid
  potential variations and so remain near their initial radius. Our
  model thus explains the flattening of central density cusps into
  kiloparsec-scale cores in small galaxies through radial outwards
  migration. As expected the reversible, adiabatic model (illustrated
  for the innermost orbit by the dashed line) does not correctly model
  the heating effect of very rapid potential variations in the inner
  parts of the halo.}\label{fig:radial-migration}
\end{figure}

The solid lines in Figure \ref{fig:radial-migration} show the
resulting mean radius $\langle r \rangle$ of orbits as a function of
time, where
\begin{equation}
\langle r \rangle = \left. \int \frac{r \, \dd r}{\sqrt{E(t)-V_{\mathrm{eff}}(r;j,t)}} \right/ \int \frac{ \dd r}{\sqrt{E(t)-V_{\mathrm{eff}}(r; j,t)}}.
\end{equation}
The values of $j$ and $E_0$ for each orbit are chosen by requiring the
initial motion to be circular at a range of different radii. As time
progresses, the orbits starting interior to $1\,\kpc$ migrate
outwards, reflecting a net gain in energy. Orbits outside this radius
are largely unaffected.  In the LT run, by contrast, no tracer
particles gain energy; those that start on circular orbits, for
instance, are predicted to remain at the same radius for the entire
run.

Figure~\ref{fig:HT-fluctuations} implies that the central baryonic
potential returns to its approximate original shape at the end of each
starburst cycle, because the gas affected by the supernovae has cooled
back into the disk (or has flowed out, replaced by fresh
gas). In the adiabatic limit, where all potential changes occur
slowly,  the final orbital parameters should return to their initial values.
Indeed by making $\Delta V_{\mathrm{eff}}$ infinitesimal and
integrating (\ref{eq:delta-E1}) one obtains
\begin{equation}
\int \sqrt{E(t) - V_{\mathrm{eff}}(r; j, t)} \,\dd r = \textrm{constant,}\label{eq:adia-gen}
\end{equation}
where again the integral is taken over the real region of the
integrand. This is the generalization of
equation~(\ref{eq:adiabatic-Efinal}), and is exactly the adiabatic
invariant derived through the action-angle approach
\citep[e.g.][]{1987gady.book.....B}. It implies that $E_{\mathrm{final}} =
E_{\mathrm{initial}}$ if the potential returns to its initial form
via a series of slow changes.

Demanding the adiabatic invariant (\ref{eq:adia-gen}) is 
constant yields the orbital migration in the `gradual outflows'
scenario.  The dashed line in Figure \ref{fig:radial-migration} shows
that the result derived in this limit is as expected: although
temporary changes in the orbital radius do occur, they do not persist
over time. This underlines the difference between our new model (where
a tracer particle picks up energy from baryons) and the older
adiabatic calculations (where the energy of a tracer particle is
conserved).

Although Figure~\ref{fig:radial-migration} shows that orbits gain energy,
it cannot be used directly to infer the final inner profile of the
dark matter. To draw conclusions about the evolution of the slope, we
evolved the energy of $\sim 90\,000$ orbits corresponding to all dark
matter particles in the halo at $z=4$. At each timestep, the full
radial probability distribution for each particle,
\begin{equation}
p(r; E, j) \propto \frac{1}{\sqrt{E-V_{\mathrm{eff}}(r; j)}}\textrm{,}
\end{equation}
was calculated numerically. The sum of the normalized probability
distributions for all particles then implies a density profile
according to
\begin{equation}
\rho_{\mathrm{model}}(r) \propto \frac{1}{r^2}\sum_i p(r; E_i, j_i)\textrm{,}\label{eq:model-density}
\end{equation}
where the sum is over all tracer particles. Time evolution of
$\rho_{\mathrm{model}}(r)$ arises from updating $V_{\mathrm{sphere}}$
and each $E_i$ at every timestep according to
equation~(\ref{eq:delta-E1}); or, for comparison, by solving
equation~(\ref{eq:adia-gen}) to derive the behaviour in the adiabatic
limit.

\begin{figure}
\includegraphics[width=0.5\textwidth]{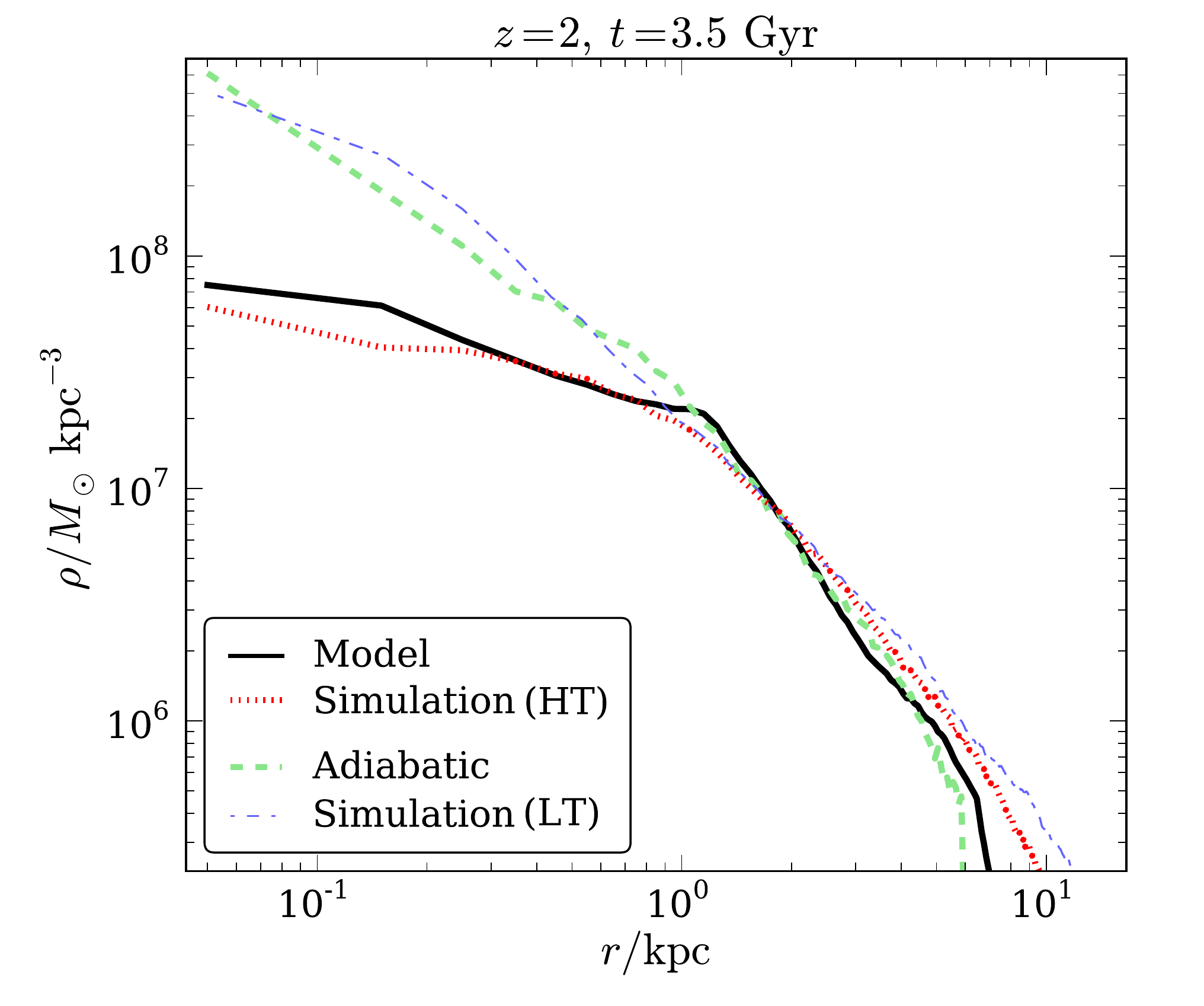}
\caption{The spherically averaged dark matter density as a function of
  radius, measured at $z=2$ when the core has formed in the HT
  simulations (thick dotted line). The solid line shows the density
  profile at this time according to our model (see text for details);
  this is seen to be in excellent agreement with the HT simulation.
  The adiabatic model (dashed line) fails to correctly model the cusp
  flattening, demonstrating the need for the improved modelling
  presented here. The LT comparison simulation (dash-dotted line) also
  remains cusped as explained in Section \ref{sec:first-sims}.
}\label{fig:final-profiles}
\end{figure}

Starting at $z=4$, the distribution function is evolved in this way to
$z=2$; the resulting density profiles are illustrated in Figure
\ref{fig:final-profiles}.  The thick solid line shows our main model
[i.e. it is derived from equation~(\ref{eq:delta-E1})], and is seen to
be in excellent agreement with the output of the simulation (dotted
line). The dotted line shows the results of modelling the baryonic
effects using the adiabatic approximation
[i.e. equation~(\ref{eq:adia-gen})]; the cusp remains, contrary to the
results of the simulation. This reaffirms that the adiabatic
approximation does not capture important aspects of the impact of
baryons on the dark matter.  Finally, the dash-dotted line shows the
profile from the LT (low star formation threshold) simulation which,
as explained in Section \ref{sec:first-sims}, retains its cusp and is
therefore in approximate agreement with the adiabatically evolved
case.

\begin{figure}
\includegraphics[width=0.5\textwidth]{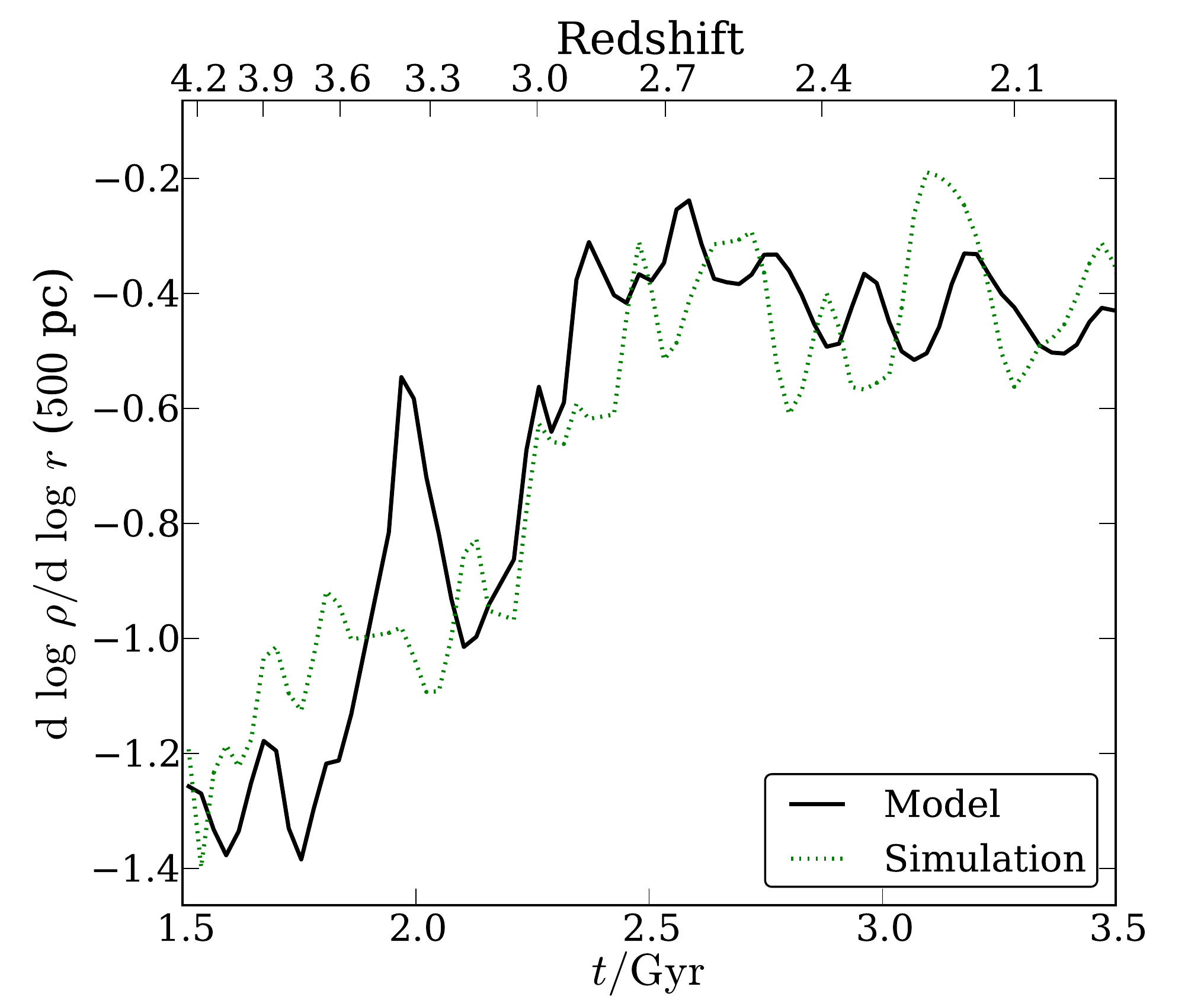}
\caption{The evolution of the logarithmic slope of the halo in the raw
  simulation data (dotted line) and according to our model (solid
  line), measured at $500\,\pc$ in both cases. Values $\lsim -1$
  indicate a conventional NFW-style cusp; shallower profiles are
  indicated by values closer to zero. The flattening of the profile in
  the simulations has previously been demonstrated to agree with
  observational data \citep{2010arXiv1011.2777O}. Our physical
  model for the origin of the flattening is in excellent agreement
  with the detailed simulation results.}\label{fig:slope-comparison}
\end{figure}

The calculation described by equation~(\ref{eq:delta-E1}) involves
calculating the particle distribution function for every intermediate
step. It is possible, therefore, to monitor the rate at which the cusp
flattens and compare it against the simulations.
Figure~\ref{fig:slope-comparison} shows the time evolution of the
measured logarithmic slope at $500\,\mathrm{pc}$ for both the model
density profile, equation~(\ref{eq:model-density}), and the simulated
density profile.  As time progresses, both the simulated and model
density profiles gradually flatten out, with the value rising from
$<-1.0$ (cusped) to $\sim -0.4$, consistent with observations
\citep{2010arXiv1011.2777O}.

To conclude, the analytic model presented here predicts
flattening of the slope at the same rate as seen in the HT simulation
(Figures \ref{fig:final-profiles} and \ref{fig:slope-comparison});
whereas the adiabatic approximation does not predict any significant
flattening (Figure \ref{fig:final-profiles}). We further verfied that
the new model did not predict a change in slope for the LT run in
which the gas density remains too small to generate significant
fluctuations in the potential. Overall, the new model alone provides a
convincing explanation for the flattening processes seen by G10.

\section{Conclusions}\label{sec:conclusions}

We have proposed a new analytic model that accounts for the flattening
of dark matter cusps into cores. Energy is transferred into dark
matter particle orbits through repeated, rapid oscillations of the
central gravitational potential. These oscillations are caused by
recurrent, concentrated bursts of star formation which induce rapid
expansion of gas through supernova feedback heating. We verified that
this process quantitatively accounts for cusp-flattening in a novel
set of simulations similar to those in G10.  The simulations include
the effects of metal-line cooling, but like those in G10 form thin
stellar disks and have a galactic star formation efficiency of only a
few per cent.  A comparison simulation (LT) with lower star formation
density threshold does not form a core, despite forming ten times as
many stars by $z=0$.  The model correctly predicts no cusp-flattening
in this case (Figure \ref{fig:slope-comparison}) confirming our
interpretation that for cores to form the supernova energy must be
injected in a concentrated region (Figure~\ref{fig:den}).

The baryons do not have to escape the system completely, but only
temporarily vacate the central regions, because the energy transfer is
inherently irreversible.  The G10 simulations do exhibit
galactic-scale outflows which remove $75\%$ of baryons by $z=0$. These
outflows have other important effects
\citep[e.g.][]{2010arXiv1010.1004B}, yet the mass involved is only
around a third of the mass involved in the central blowout-recollapse
cycle. The relation between the galactic outflows and the local
feedback will be the focus of future work.

The picture is related to the more established view that removal of
baryons through galactic superwinds could cause the central dark
matter profile to flatten
\citep[e.g.][]{1996MNRAS.283L..72N,2002MNRAS.333..299G,
  2005MNRAS.356..107R}.  However, our model confirms that extreme,
violent mass-loss events are not necessary, as suggested by
\citep{2006Natur.442..539M,2008Sci...319..174M}.  This is important
because the more moderate heating events allow retention of baryons
and the formation of a thin stellar disk.

Dynamical friction from infalling baryonic clumps
\citep{2001ApJ...560..636E,mo04,2009ApJ...702.1250R,2010ApJ...725.1707G}
does not appear to play a dominant role in our simulations. {\revised
  In the LT simulations, no cores form; therefore effects of dynamical
  friction are ruled out except from the densest clumps in HT.  In the
  HT simulations, any dense infalling clumps are typically disrupted
  long before they reach the inner kiloparsec.  To test the effect of
  dynamical friction, one would first need to remove the explosive
  events associated with feedback. However as the feedback energy is
  decreased, dense clumps start to pile up in galaxies until time
  integration becomes computationally unfeasible, at the same time
  creating an unrealistic dense central bulge. Additionally we do not
  have the resolution to produce star clusters
  \citep{2010ApJ...725.1707G} which could be more robust to disruption
  than gas clumps. As a result we are not ruling out dynamical
  friction as an agent for weakening cusps except for the particular
  simulations in use here.}

The DG1 case (which undergoes major mergers at $z=3$ and $z=1$; see
Section \ref{sec:first-sims}) has a very similar cusp-flattening
history to the galaxy described in this work. The mergers themselves
have no measurable effect on the halo slope and, as expected through
arguments based on Liouville's theorem, cores remain present after
the merger \citep{2006ApJ...641..647K}. In Governato et al (in prep.)
we show that cores form in the large majority of dwarf galaxies,
irrespective of their assembly history.

The model is reminiscent of the violent relaxation envisioned by
\cite{1967MNRAS.136..101L}, although our analysis does not attempt to
generate the gravitational potential self-consistently.  It would thus
be desirable in the future to work the microphysics into a broader,
analytical description of the evolution of a self-consistent dark
matter distribution function.  We would further like to investigate
the importance of departures from exact spherical symmetry. The
simulated supernova explosions are rarely exactly on the axis of the
disk, so even axisymmetry is violated. This will lead to the
non-conservation of angular momentum, which must be understood before
we can investigate the impact of the process on the anisotropy of the
orbits \citep{2006ApJ...649..591T}.

Stars, like dark matter particles, are collisionless and therefore
should be subject to the same migratory processes outlined here.
Stars forming near the centre of the HT galaxies indeed migrate
outwards;  it could be natural in this context that the scale-length
of the dark matter cores and the stellar disks are approximately
equal, an observational relation noted by
\cite{2009Natur.461..627G}. Indeed the scale-lengths of both stellar
disk and dark matter core of the dwarf galaxies discussed here are
approximately 1 kpc (G10; \citeauthor{2011ApJ...728...51B} 2011).
To make this link convincing will require a more systematic study of
scaling with mass (Brooks et al in prep).  


{\revised While the analytic model we have described is independent of
  the detailed gas dynamics, to obtain results the simulated
  hydrodynamics are used as an input. The cusp-flattening effect is
  therefore only achieved in reality if the rapid gas motions
  predicted by the SPH code are reproduced. Mesh-based codes
  \citep{2002A&A...385..337T,2004astro.ph..3044O,2010MNRAS.401..791S}
  highlight potential shortcomings of traditional SPH such as its poor
  handling of instabilities related to sharp density contrasts
  \citep{2007MNRAS.380..963A,2011arXiv1109.4413B}. In future we intend
  to address the sensitivity of our results to these inaccuracies
  through comparison with alternative codes and use of forthcoming
  improvements to the {\it Gasoline} SPH engine that reduce artificial
  surface tension \cite[see also][]{2011arXiv1111.6985R}. Through
  direct comparison of various codes \cite{2011arXiv1112.0315S}
  conclude that, for most practical purposes, the choice of sub-grid
  model approximation (via which supernova energy is coupled to the
  gas) is more critical than the numerical technique. It will be of
  interest to determine whether other feedback mechanisms and
  numerical methods can reproduce our results. Indeed recently
  \cite{2011arXiv1112.2752M} reported the formation of dark matter
  cores in AMR simulations through gas fluctuations very similar to
  ours, but driven by AGN activity in clusters.

  We are currently investigating how the cores scale for $10^9\,
  \mathrm{M}_{\odot} < M_{\mathrm{vir}} < 10^{12} \,
  \mathrm{M}_{\odot}$ (Governato et al, in prep). At higher masses the
  results will depend on a detailed interplay between the deepening
  dark matter potential, increased star formation rates, and the
  nature of AGN feedback \citep{2011arXiv1112.2752M}. }  The challenge
of running suitable simulations to tackle the most massive systems at
sufficient resolution is formidable, but one that we hope to tackle in
due course.


\vspace*{1cm}

\section*{Acknowledgements}

We thank the anonymous referee for helpful comments. AP thanks Steven
Gratton for many helpful discussions and James Wadsley, Justin Read,
Jorge Pe\~{n}arrubia, Alyson Brooks, Fergus Simpson, Hiranya Peiris,
Gary Mamon and Max Pettini for comments on a draft version of the
paper.  FG acknowledges support from a NSF grant AST-0607819 and NASA
ATP NNX08AG84G.

The analysis was performed using the pynbody package
(\texttt{http://code.google.com/p/pynbody}). Simulations and analysis
were performed using the Darwin Supercomputer of the University of
Cambridge High Performance Computing Service
(\texttt{http://www.hpc.cam.ac.uk}), provided by Dell Inc. using
Strategic Research Infrastructure Funding from the Higher Education
Funding Council for England.

\bibliographystyle{apj} \bibliography{../refs.bib}

\end{document}